\documentclass{iau} 
\usepackage{graphicx}

\usepackage{natbib}

\title[Galaxy Mergers and the Main Sequence] 
{Deep Learning for Galaxy Mergers in the Galaxy Main Sequence}

\author[William J. Pearson et al.]   
{William J. Pearson$^{1,2}$,
 Lingyu Wang$^{1,2}$,
 James Trayford$^3$,
 Carlo E. Petrillo$^2$,
 \and  Floris F.S. van der Tak$^{1,2}$}

\affiliation{$^1$SRON Netherlands Institute for Space Research\\ Landleven 12, 9747 AD, Groningen, The Netherlands\\ email: {\tt w.j.pearson@sron.nl} \\[\affilskip]
$^2$Kapteyn Astronomical Institute, University of Groningen\\ Postbus 800, 9700 AV Groningen, The Netherlands\\
[\affilskip]
$^3$Leiden Observatory, Leiden University, P.O. Box 9513, 2300 RA Leiden, The Netherlands}

\pubyear{2018}
\volume{341}
\jname{PanModel2018 : Challenges in Panchromatic Galaxy Modelling with Next Generation Facilities}
\editors{M. Boquien, E. Lusso, C. Gruppioni \& P. Tissera}
\begin{document}

\maketitle

\begin{abstract}
Starburst galaxies are often found to be the result of galaxy mergers. As a result, galaxy mergers are often believed to lie above the galaxy main sequence: the tight correlation between stellar mass and star formation rate. Here, we aim to test this claim.

Deep learning techniques are applied to images from the Sloan Digital Sky Survey to provide visual-like classifications for over 340\,000 objects between redshifts of 0.005 and 0.1. The aim of this classification is to split the galaxy population into merger and non-merger systems and we are currently achieving an accuracy of 92.5\%.  Stellar masses and star formation rates are also estimated using panchromatic data for the entire galaxy population. With these preliminary data, the mergers are placed onto the full galaxy main sequence, where we find that merging systems lie across the entire star formation rate - stellar mass plane.
\end{abstract}

\firstsection 
\section{Introduction}
Galaxy mergers are highly important for fully understanding how galaxies form and evolve, underpinning hierarchical growth models. Also, galaxy mergers are believed to be the driving force behind some of the brightest infrared objects known \citep{1996ARA&A..34..749S}. With bright infrared emission often comes high star formation rates (SFRs), hence the argument that most mergers go through a starburst phase \citep[e.g.][]{2005ASSL..329..143S}.

More recent studies have begun to show that merging systems may not always cause starbursts. \citet{2015MNRAS.454.1742K} have found that the increase in SFR in merging galaxies is at most a factor of two in the local universe, with the majority of galaxies showing no evidence of an enhance SFR or showing evidence of merger driven quenching. However, mergers can still cause starbursts with the majority of starbursts being caused by galaxy mergers \citep{2014ApJ...789L..16L, 2015ApJ...807L..16K, 2017A&A...607A..70C}.

The main sequence of star forming galaxies (MS) is a tight correlation between the stellar mass (M$_{\star}$) and SFR of star forming galaxies \citep[e.g.][]{2004MNRAS.351.1151B, 2007ApJ...660L..43N, 2014ApJS..214...15S}. The MS has been shown to be present out to at least $z = 6$, although the presence, or not, of a high mass turnover is still debated \citep[e.g.][]{2016ApJ...817..118T, 2018A&A...615A.146P}. Whatever the form it takes, the majority of galaxies are found to lie on the MS, with high SFR objects and starbursts above and quenched or quiescent objects below. Thus, if all merging galaxies are starbursts, all merger galaxies should lie above the MS, otherwise they will be scattered around the SFR-M$_{\star}$ plane.

Deep learning neural networks are a rapidly growing technique that can be used to classify galaxies. These techniques use layers of non-linear mathematical functions that are inspired by the functioning of biological networks of neurons. Once properly trained, a neural network can classify thousands of galaxies in a fraction of the time it would take a human, or team of humans, to perform the same task. Deep neural networks are increasingly used in astronomy to perform tasks such as determining galaxy morphology \citep[e.g.][]{2015ApJS..221....8H}, detecting lensed systems \citep[e.g.][]{2017MNRAS.472.1129P} and detecting galaxy mergers \citep[e.g.][]{2018MNRAS.479..415A}.

\section{Stellar Masses and Star Formation Rates}
To populate the SFR-M$_{\star}$ plane, we need SFRs and M$_{\star}$s for the objects. For the SDSS data release 7 galaxies with spectroscopic redshifts between 0.005 and 0.1, the five SDSS bands (ugriz) were run through CIGALE \citep{2009A&A...507.1793N, 2018arXiv181103094B} using the same configuration as \citet{2018A&A...615A.146P} but removing the active galactic nuclei component to generate M$_{\star}$, and U-V and V-J colour estimates. The SFR estimates were derived from the SDSS H$\alpha$ spectral lines but still need to be corrected for extinction.

We intend to improve the M$_{\star}$ and SFR estimates, from CIGALE, over the region of SDSS that coincides with the Galaxy And Mass Assembly survey \citep[GAMA;][]{2009A&G....50e..12D}. The GAMA region contains data from the ultra-violet (UV) to the far-infrared (FIR), allowing better constraints on the stellar mass and the thermal dust emission than just using the five SDSS bands. We will improve the FIR data in the GAMA fields by following \citet{2017A&A...603A.102P} to de-blend the \textit{Herschel} SPIRE \citep{2010A&A...518L...3G} data. This involves running CIGALE with the UV to near-infrared data to predict the flux density in the three SPIRE bands and using these predictions as a prior in XID+ \citep{2017MNRAS.464..885H} to de-blend the SPIRE images. With more reliable \textit{Herschel} SPIRE data from this technique, we can generate more reliable SFR estimates.

\section{Identifying Mergers}
To identify merging systems, a convolutional neural network (CNN) was employed. This type of neural network uses a series of kernels to identify features in the input image and create classifications, in this case if the galaxy is merging or not. The network used is based on the \citet{2015MNRAS.450.1441D} architecture with some changes to the parameters of the four convolutional layers and two fully connected layers, as well as changing the input to be a 64$\times$64 pixel, 3 colour channel image. To train the network, the 3003 Sloan Digital Sky Survey \citep[SDSS;][]{2000AJ....120.1579Y} objects classified as merging by \citet{2010MNRAS.401.1552D, 2010MNRAS.401.1043D} were used along with 3003 randomly selected galaxies that had Galaxy Zoo merger classification rate of less than 0.2 and fell within the same redshift range \citep{2011MNRAS.410..166L, 2018MNRAS.479..415A}. After training, this resulted in an accuracy (percentage of objects that are correctly classified) of 92.5\%. The network was then applied to over 340\,000 SDSS galaxies with redshifts between $z = 0.005$ and $z = 0.1$, categorising them in approximately two hours. Approximately 19\% of these objects are preliminarily classified as being merging systems, twice the merger fraction seen in other observational works \citep{2016ApJ...830...89M}.

\section{The Main Sequence of Starforming Galaxies}
\begin{figure}
	\centering
	\includegraphics[width=0.7\textwidth]{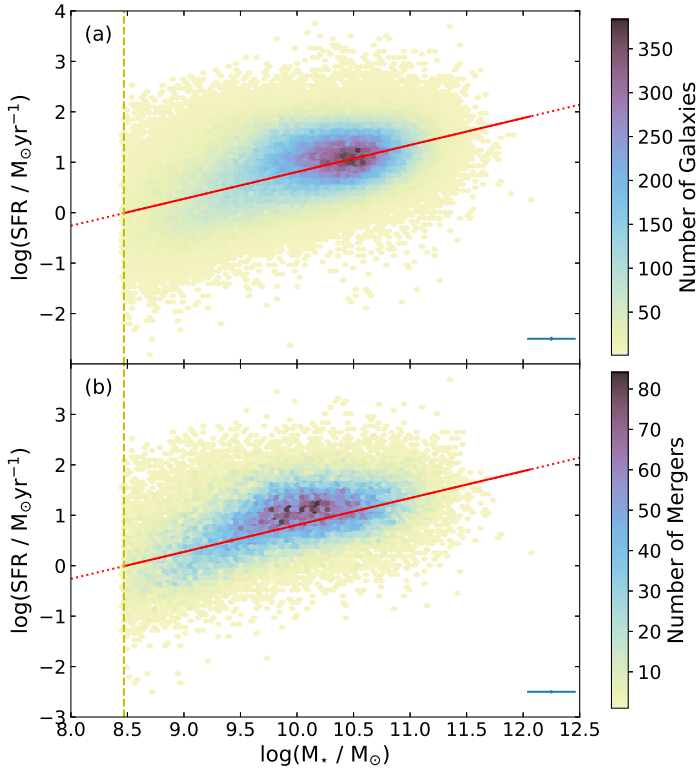}
	\caption{M$_{\star}$-SFR plane as a number density plot from low (light) to high (dark) with (a) all SDSS objects with detections (at any confidence level) of H$\alpha$ and H$\beta$ lines above the mass limit (vertical dashed line) and (b) only the objects from (a) that are identified as mergers. The solid line is the main sequence trend, with extrapolations to higher and lower mass as a dotted line. As can be seen, the merging systems closely follow the entire galaxy population, perhaps with a slight rise, and are not preferentially starbursting or quenched.}
	\label{MS-fig}
\end{figure}

To find the trend of the MS for this data set, we follow \citet{2018A&A...615A.146P}. The quiescent galaxies are split off using a UVJ colour cut and the 90\% mass completeness limit determined empirically by following \citet{2010A&A...523A..13P} and using predicted Ks-band magnitudes from CIGALE. A simple power law is fitted to the remaining objects using a Markov Chain Monte Carlo routine to simultaneously fit the slope, normalisation and scatter of the MS, taking into account the observational errors of both the SFR and M$_{\star}$. The resulting trend is plotted in Fig. \ref{MS-fig} as a solid line.

To determine if all mergers are starbursts, we present two plots. Fig. \ref{MS-fig}a shows the SFR-M$_{\star}$ plane populated with all SDSS galaxies that have detections (at any confidence level) of H$\alpha$ and H$\beta$ lines above our mass completeness limit. As can be seen the MS trend runs through the high number density regions. In Fig. \ref{MS-fig}b we also plot the SFR-M$_{\star}$ plane populated only objects in Fig. \ref{MS-fig}a that our CNN identifies as mergers. As can be seen, there is no excess of galaxies above the MS, as would be expected if every merger caused a starburst at the time of observation. Equally, there is no clumping of objects below the MS, showing no evidence that all mergers trigger quenching. Indeed, the merging systems follow the distribution of the entire galaxy population, perhaps with a slight rise in SFR for the merger population. With this preliminary result, there is no evidence that all merging systems are starbursts.

\section{Ongoing Work}
We are currently working to train a neural network with simulated SDSS images from the EAGLE simulation \citep{2015MNRAS.446..521S}. This network will be applied to the observed SDSS images, allowing us to test for possible biases and incompleteness in human classification: galaxies that are classed by humans as merger or non-merger but are given the opposite class by the EAGLE trained network. This also has applications for the upcoming large surveys, such as \textit{Euclid} \citep{2011arXiv1110.3193L} and the Large Synoptic Survey Telescope \citep{2009arXiv0912.0201L}. If this technique proves to function as well as, or better than, an observationally trained network, it will reduce the need to first create a sample of galaxies for humans to classify for training a network. The simulation trained network can immediately be applied to the data from the large surveys and create data sets for science quickly.

Similarly, the simulated images can be passed through the observation trained network. This will allow us to check that the simulation data are representative of the observed universe. Objects from simulations that are classified as mergers but the observation trained network classifies as non-mergers can be identified. Understanding why these objects are rejected as mergers by observations could provide insight into shortcomings of the simulations and inform future work simulating our Universe.

\pagebreak

\begin{discussion}

\discuss{Fabio Fontanot}{What is the fraction of merging galaxies as a function of stellar mass?}

\discuss{William J. Pearson}{This is not something we have looked at yet but it would be interesting to see. (A quick look after the talk shows that the merger fraction decreases as stellar mass increases).}

\discuss{Minju Lee}{It may be worth to separate the sample into two groups of gas-rich and gas-poor to test if gas-rich mergers cause SB while gas-poor don't.}

\discuss{William J. Pearson}{Yes, it would indeed be interesting to split by gas-rich/poor to see where they lie. I have not done this yet but would be interesting.}
\end{discussion}

\end{document}